\documentclass[fleqn,twoside]{article}
\usepackage[headings]{espcrc2}
\usepackage{amsmath}
\usepackage{amssymb}
\usepackage{graphicx}

\newcommand{\sla}[1]{{/\hskip-0.55em{}#1}{}}

\title{%
Automated Calculation Scheme for $\alpha^n$ Contributions of QED 
to Lepton $g-2$\thanks{%
Talk presented by T.~Aoyama at the 7th International Symposium on Radiative Corrections (RADCOR 2005), Shonan Village, Japan, October 2-7, 2005.}}

\author{
T.~Aoyama\address[RIKEN]{%
Theoretical Physics Laboratory, RIKEN, Wako, Saitama, Japan 351-0198}, 
M.~Hayakawa\addressmark[RIKEN],
T.~Kinoshita\address[Cornell]{%
Laboratory for Elementary Particle Physics, Cornell University, Ithaca, New York 14853, U.S.A.},
and 
M.~Nio\addressmark[RIKEN]
}
       
\runtitle{%
Automated Calculation Scheme for $\alpha^n$ Contributions of QED 
to Lepton $g-2$}
\runauthor{%
T.~Aoyama, M.~Hayakawa, T.~Kinoshita, and M.~Nio}

\begin{document}
\begin{abstract}
This article reports an automated approach to the evaluation of 
higher-order terms of QED perturbation to anomalous magnetic 
moments of charged leptons by numerical means. 
We apply this approach to tenth-order correction due to 
a particular subcollection of Feynman diagrams, 
which have no virtual lepton loops. 
This set of diagrams is distinctive in that it grows factorially 
in number as the order increases, 
and also each of the diagrams holds quite a large number of 
subtraction terms to be treated along renormalization procedure. 
Thus some automated scheme has long been required to evaluate 
correctly this class of diagrams. 
We developed a fast algorithm and an implementation which automates 
necessary steps to generate from the representation of 
each Feynman diagram the FORTRAN codes for numerical integration. 
Currently those diagrams of tenth order are being evaluated.
\end{abstract}

\maketitle

\section{Introduction}
\label{sec:introduction}
The anomalous magnetic moment of electron, 
also called the electron $g\!-\!2$, 
is one of the fundamental quantities of particle physics. 
Since its discovery in 1947 \cite{kusch} 
it has been measured with steadily 
increasing precision \cite{rich,VanDyck:1987ay}. 
The best values of $g\!-\!2$ of the electron and the positron 
available in the literature \cite{VanDyck:1987ay} 
\begin{equation}
\begin{aligned}
        a_{e^-}({\rm exp}) 
        &=
        1\ 159\ 652\ 188.4\ (4.3) \times 10^{-12}
        \,, \\
        a_{e^+}({\rm exp}) 
        &= 
        1\ 159\ 652\ 187.9\ (4.3) \times 10^{-12}
\label{eq:expValue} 
\end{aligned}
\end{equation}
were obtained by the Penning trap experiment. 

The electron $g\!-\!2$ (denoted by $a_e = \frac{1}{2}(g-2)$) 
is explained almost entirely by the electromagnetic interaction 
between electron and photon alone, 
and thus it has provided the most stringent test of QED. 
An important by-product of the study of the electron $g\!-\!2$ is 
that currently the most precise estimation 
of the fine structure constant $\alpha$ 
can be obtained by combining the measurement and the theory of $a_e$, 
which yields \cite{kn2} 
\[
	\alpha^{-1}(a_e) = 137.035~998~834~(12)(31)(502),
\label{eq:alpha_ae}
\]
where the uncertainties 12 and 31 are due to $\alpha^4$ and $\alpha^5$ 
terms, and 502 comes from the experiment (\ref{eq:expValue}) . 

At present a new experiment on the electron $g\!-\!2$ is being carried out 
by a Harvard group, which will reduce the measurement 
uncertainty substantially \cite{Gabrielse}. 
It will enable us to test the validity of QED to a very high degree 
and to determine $\alpha$ to an unprecedented precision of 
$7\times 10^{-10}$ or better. 
Of course such a feat requires availability 
of the theoretical calculation of matching precision. 
The present issue is that 
the uncertainty due to $\alpha^5$ term may become 
major source of systematic errors. 
Therefore 
a reliable theoretical estimation of $\alpha^5$ corrections 
is urgently required. 

From the viewpoint of obtaining the $\alpha^5$ corrections 
the numerical integration approach is the only practical choice at 
present. We employ the numerical evaluation scheme developed 
by one of the authors (T.~K.) and Cvitanovi\'c 
\cite{Cvitanovic-Kinoshita,Kinoshita_book}. 

The contribution to the $\alpha^5$ term of the electron $g\!-\!2$ 
comes from 12672 vertex-type Feynman diagrams, which can be 
categorized into 6 sets according to their structures and 
classified further into 32 gauge-invariant sets 
\cite{kn3,kn-radcor05}. 
None of 32 sets is dominant so that all must be evaluated. 
A particularly difficult one is Set V, a huge set consisting 
of 6354 vertex diagrams, all of which have radiative corrections 
only due to virtual photons. 
We begin by exploiting the equation derived from the Ward-Takahashi 
identity, 
\[
        \Lambda^\nu(p,q) 
        \simeq -q^\mu \biggl[
                \frac{\partial\Lambda_\mu(p,q)}{\partial q_\nu}
        \biggr]_{q=0}
        -\frac{\partial\,\Sigma(p)}{\partial p_\nu} \,,
\label{eq:WI}
\]
which relates the sum of 9 vertex diagrams to one self-energy-like 
diagram. 
This relation together with the time-reversal symmetry of QED 
enables us to reduce the number of independent integrals of 
Set V diagrams drastically from 6354 to 389. 
 
The difficulty of Set V also stems from the fact that 
many of them have very large number of UV and IR divergences, 
\textit{e.g.} maximally 47 UV counterterms are required for 
some diagrams and many of the others require more than 20 
UV counterterms. 
This makes the previous approach highly impractical since 
it runs into an extremely sever logistic problem. 
Toward this difficult problem we present our solution 
by an automated scheme for code generation, which enables 
us to obtain renormalized integrals for all diagrams of Set V 
that are ready to be integrated by numerical means \cite{auto}.

\section{Diagrams without lepton loops}
\label{sec:setv}
\begin{figure}
\begin{center}
\includegraphics[scale=.75]{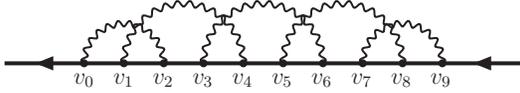}
\vskip -6ex
\end{center}
\caption{A typical diagram of Set V, which is 
represented as $\{(0,2)(1,4)(3,6)(5,8)(7,9)\}$.
\label{fig:diagram}}
\end{figure}

Set V diagrams summed by W-T identity are one-particle-irreducible 
(1PI) self-energy-like diagrams of $2n$th order which have 
no closed lepton loops. 
A typical diagram is shown in Figure~\ref{fig:diagram}. 
They are given by a path consisting of all lepton lines 
emanating from the incoming lepton and terminating at the 
outgoing lepton. 
$2n$ interaction vertices lie on the path, which are labeled 
by $v_0, v_1, \dots, v_{2n\!-\!1}$ from the left to the right 
in our conventions. 
A photon line labeled by $k$ connects the vertices 
$v_{i_k}$ and $v_{j_k}$ at its ends. 
Therefore a diagram is uniquely specified by a single-line 
expression $\{(v_{i_k},v_{j_k}),\dots\}$ which represents 
the pattern of pairings how photon lines connect the 
vertices if we adopt a proper ordering of pairs. 

All Set V diagrams are generated systematically by finding 
every possible pattern of pairings that satisfies 1PI condition, 
which reduces to a matter of simple combinatorics. 
Due to their simple structure the graph-theoretical notions, 
especially the independent set of closed paths on the diagram, 
are easily identified. It is crucial in constructing the 
integrand. 
Another significant feature is that UV-divergent subdiagrams 
are easily found as single segments on the lepton path. 
The UV-divergent structure of the diagram will be identified 
exactly by \textit{forests} 
constructed by the combinations of the subdiagrams. 
Those properties are quite efficiently handled in an automated 
manner.

\section{Numerical integration formalism}
\label{sec:formalism}
The anomalous magnetic moment $a_e$ is given by the static limit of 
the magnetic form factor $F_2(q^2)\bigr|_{q^2\to 0}$. 
The contribution from a $2n$th order Feynman diagram $\mathcal{G}$ 
is expressed by an integral over $n$ loop momenta $\{k_r\}$:
\[
\begin{aligned}
	\frac{1}{i}\Sigma_{\mathcal{G}} 
	&= 
	(ie)^{2n}
	\left[\prod_{r=1}^{n}\!\int\!\!\frac{d^4k_r}{(2\pi)^4}\right]\, 
\\ &
	\gamma^{\mu_1} \frac{i}{{\sla{p}}_1\!-\!m} 
	\cdots 
	\frac{i}{{\sla{p}}_{2n\!-\!1}\!-\!m} \gamma^{\mu_{2n}} 
	\prod_{r=1}^{n} \frac{-ig_{\mu_i\mu_j}}{k_r^{\ 2}} \\
\end{aligned}
\]
where $\prod_{r=1}^{n} \frac{-ig_{\mu_i\mu_j}}{k_r^{\ 2}}$ 
is a diagram-specific product specified by the pairing pattern 
in the case of Set V diagrams. 
The momentum integration is exactly carried out and 
the amplitude is converted into an integral over 13-dimensional 
Feynman parameter space: 
\[
\begin{aligned}
	\frac{1}{i}\Sigma_{\mathcal{G}} 
	&= 
	\left(\frac{\alpha}{\pi}\right)^{n}\!
	\frac{1}{4^n}\Gamma(n\!-\!1)
	\int(dz)_{\mathcal{G}}\,
\\ &
	\left[
        \frac{F_0(B_{ij},A_j,C_{ij})}{U^2 V^{n-1}} 
        + \frac{F_1(B_{ij},A_j,C_{ij})}{U^3 V^{n-2}} + \cdots \right].
\end{aligned}
\]
The result is expressed symbolically as a function of quantities 
$U, B_{ij}, A_j, C_{ij}$ and $V$ (called as building blocks), 
which are homogeneous polynomials of Feynman parameters. 
The explicit forms of $U$ and $B_{ij}$ are determined by the 
underlying topological structure of the Feynman diagram $\mathcal{G}$ 
\cite{Nakanishi:1971}. 
$A_j, C_{ij}$ and $V$ also reflect the structure of the diagram 
and are obtained from $U$ and $B_{ij}$. 

The amplitude constructed above is divergent in general, 
and the divergences must be removed before carrying out 
the integration numerically. 
We adopt the subtractive on-shell renormalization. 
In order to perform renormalization numerically 
we employ a strategy in which 
we prepare the subtraction terms as integrals over the same 
domain of the original unrenormalized amplitude, and 
carry out point-wise subtraction so that the singularities 
of the original integrand are canceled point-by-point on the 
Feynman parameter space. 

To achieve this we adopt the following intermediate 
renormalization scheme. 
We first subtract the UV-divergent part of the unrenormalized 
integrand, which is identified by simple power-counting rules. 
This procedure is formulated as \textit{K}-operation. 
By construction the subtraction term factorizes exactly into 
UV-divergent part of the renormalization constant $L_{\mathcal{S}}$ 
associated to the UV-divergent subdiagram $\mathcal{S}$ and 
the lower-order $g\!-\!2$ term $M_{\mathcal{G}/\mathcal{S}}$. 
The factorization property is significant in the present formulation 
in that it allows successive application of \textit{K}-operations. 

The whole subtraction terms are given by Zimmermann's forest formula 
\cite{Zimmermann:1969jj}.
Each source of UV-divergence is related to a \textit{forest}, 
a set of non-overlapping UV-divergent subdiagrams. 
The subtraction term associated to a forest is constructed by 
applying \textit{K}-operations to the unrenormalized integrand 
successively for the subdiagrams in a proper order. 
The advantage of the forest approach is that it is readily 
translatable into code generation. 
The forests are given by the combination of the subdiagrams, so the 
complete identification of UV-divergent parts are obtained by 
purely combinatorial procedure. 
Thus it is quite efficiently implemented in terms of forests, 
which enables us to obtain fully UV-renormalized amplitude of 
the Feynman diagram $\mathcal{G}$. 

The diagram may have IR-divergences when it contains self-energy 
subdiagrams. 
The subtraction of IR divergences is achieved by 
\textit{I}-operation 
in a similar manner as that of UV divergences. 

Counterterms thus constructed can be identified with only the 
divergent part of the renormalization constants so that the 
result of the above step is not fully equivalent to the 
standard on-shell renormalization. 
To complete the calculation the difference between 
the full renormalization and the intermediate renormalization 
must therefore be evaluated 
by summing up all subtraction terms. 
This step is called the residual renormalization, 
which is also much complicated for the higher-order calculations 
and is to be automated in the future work.

\section{Automated flow of calculation}
\label{sec:autoflow}
\begin{figure*}[t]
\begin{center}
\includegraphics[scale=0.7]{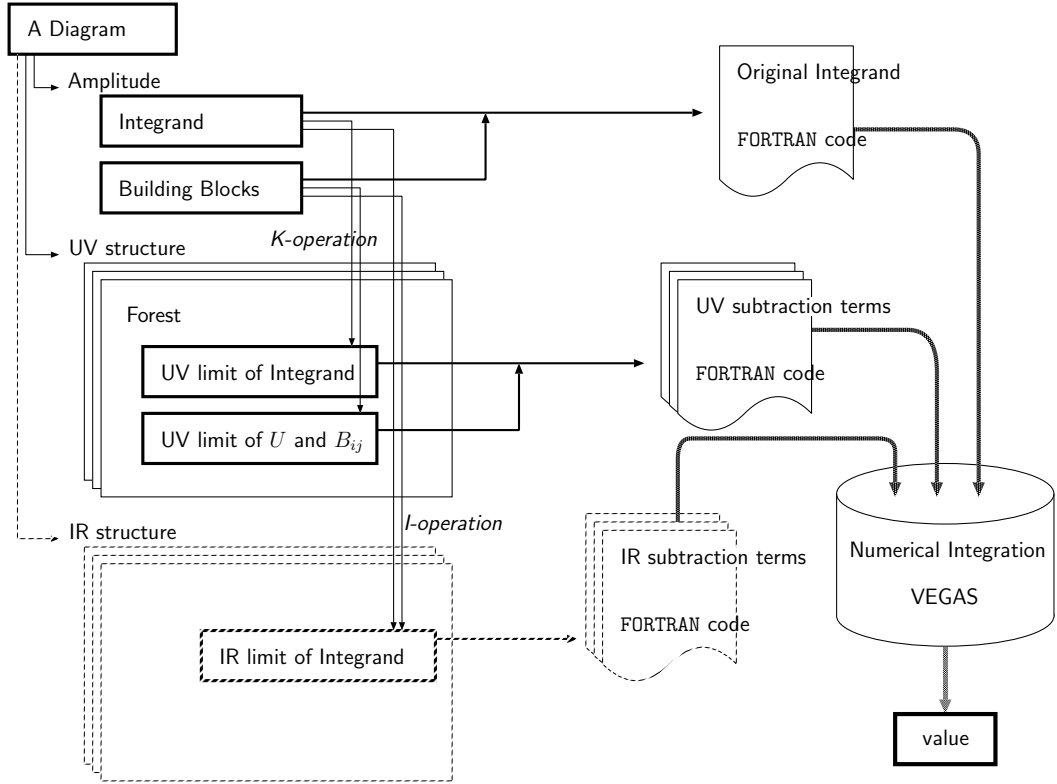}
\vskip -4ex
\end{center}
\caption{%
Flow of process to generate the numerical integration codes 
from the diagram representation.
\label{fig:flow}}
\end{figure*}
The flow of the process to generate 
the numerical integration code for evaluating an individual diagram 
is schematically depicted as Figure~\ref{fig:flow}. 

The information of a diagram is provided 
in a single-line representation indicated by the rectangular box 
at the upper-left corner. 
It enables us to construct the unrenormalized amplitude 
determined directly by the pairing pattern of the diagram, 
which is converted into the form of Feynman parametric integrals 
in terms of building blocks by analytically performing 
the integration over the loop momenta 
with the help of algebraic manipulation program, 
FORM \cite{vermaseren}. 

The explicit form of the build blocks 
are determined from the underlying topological structure. 
They are obtained by referring to the fundamental set of 
closed paths on the diagram, which can be extracted from 
the diagram representation. 
This step also incorporates algebraic manipulations 
performed by MAPLE. 

The UV divergences of the diagram are identified as 
Zimmermann's forests, which are obtained by the 
non-overlapping combinations of UV-divergent subdiagrams 
once the complete set of those subdiagrams are found 
from the diagram representation. 
The subtraction operation is achieved by \textit{K}-operation, 
in which simple power-counting rules are applied 
to the original unrenormalized integrand and the building blocks 
to result in the proper form of subtraction term. 
A subtraction term corresponding to a particular forest is 
constructed by the successive application of \textit{K}-operations, 
each of which is associated to the subdiagram in the forest. 

The IR divergences remaining in the individual diagram should also 
be subtracted away by \textit{I}-operations in a similar manner 
as those of UV divergences, 
though this subject is not covered in the present article. 

Finally, the (intermediate-) renormalized amplitude constructed from 
the original amplitude and the set of subtraction terms 
is turned into a FORTRAN code. 
It is readily processed by the numerical integration system 
such as VEGAS \cite{lepage}, 
an adaptive Monte-Carlo integration routine.

\section{Current status and concluding remarks}
\label{sec:conclusion}
In this article we presented an automated scheme of code generation 
for evaluating higher-order QED corrections of the electron anomalous 
magnetic moment by numerical means. 
We constructed an algorithm and concrete procedure to obtain 
UV-renormalized amplitudes for a particular set of diagrams without 
lepton loops in an automated manner. 
It should be noted that the scheme itself is applicable to 
an arbitrary order of diagrams of this particular type, 
though our practical concern is to evaluate the tenth-order corrections. 

We obtained a program to list up 
all the topologically distinct diagrams, and to identify the 
UV divergent structure of each diagram in terms of forest structures. 
We implemented our automated procedure as a set of Perl programs 
with the help of symbolic manipulation systems, FORM and MAPLE. 
From a single-line representation of a diagram it generates 
numerical integration codes in FORTRAN, which are ready to 
be processed by VEGAS, an adaptive Monte-Carlo integration 
routine. 

The programs have been tested for lower-order diagrams and 
confirmed that they reproduce the codes for the sixth-order 
and eighth-order diagrams previously constructed. 
They are now being applied to tenth-order diagrams. 
At present, the diagrams which have only vertex renormalization 
were processed and test runs were performed. 
Those diagrams corresponds to 2232 vertex diagrams among 
6354 Set V diagrams of tenth-order.
Crude evaluation showed no sign of divergent behavior, 
which confirms that our scheme is working well as expected. 
They are currently put to production runs. 

Typical size of the numerical integration code of a diagram 
amounts to about 80,000 lines of FORTRAN program. 
It takes 10--20 minutes to generate the code on an ordinary PC. 
The running time to evaluate $10^6$ samplings times $20$ iterations 
takes about 5--7 hours on 32 CPU cluster. 
It seems that the estimated computational cost required to 
accomplish the whole numerical integration in a few percent 
of uncertainty remain within the manageable scale 
with up-to-date computational facilities. 

The remaining 4122 diagrams have not only UV-divergent 
self-energy subdiagrams but also IR divergences. 
The simplest way to deal with the IR problem is 
to give a small mass $\lambda$ to photons, which requires 
no further work on the automating code, and is being pursued 
as a first step. 
To obtain a result independent of $\lambda$ it is necessary 
to incorporate IR subtraction terms in a manner similar to 
that of UV counterterms, which is left as our future issue.


\end{document}